\documentstyle[epsf,a4,12pt]{article}

\begin{document}

\begin{titlepage}

\vspace*{2.truecm}

\centerline{\Large \bf The Critical Exponent $\theta^\prime$ in Spin Glasses}

\vskip 2.0truecm
\centerline{\bf H.J. Luo, L. Sch\"ulke, and B. Zheng$^*$}
\vskip 0.2truecm

\vskip 0.2truecm
\centerline{Universit\"at -- GH Siegen, D -- 57068 Siegen, 
Germany}

\centerline{$^*$Universit\"at -- Halle, D -- 06099 Halle, Germany}

\vskip 2.5truecm

\abstract{ Short-time dynamic scaling behavior of the
3D $\pm J$ Ising spin glass is studied by Monte Carlo methods.
Starting the replicas with independent 
initial configurations with a small pseudo magnetization,
the dynamic evolution of the overlap $q(t)$ between two replicas
is measured.
The initial increase of the overlap $q(t)$ is observed and
the corresponding exponent $ \theta ^ \prime $ is obtained.
From the scaling relation $ \lambda =d/z-\theta ^ \prime$,
the dynamic exponent $z$ is estimated.
 
}

\vspace{0.3cm}


\end{titlepage}

For systems at criticality, the spatial correlation length
and the correlation time are divergent. 
Relevant physical observables present singular behavior,
 characterized
by a power law and scaling.
Traditionally, it was believed that universal scaling behavior only 
emerges in equilibrium or in 
the long-time regime of the dynamic evolution. 
However, recent works have revealed that the scaling behavior
already emerges at the {\it macroscopic} early stage of the
dynamic evolution, after a microscopic time scale $t_{mic}$ \cite{jan89}. 
One typical example is that a magnetic system initially at high temperature
with a small magnetization is suddenly quenched to the critical
temperature (without any external fields)
 and then released to the dynamic evolution of model A 
\cite{hoh77}.
A new critical exponent $x_0$ has been introduced to describe the 
scaling dimension of the initial magnetization. It was found that
at the early stage of the dynamic evolution, the magnetization 
obeys a power law 
\begin{equation}
M(t)\sim m_0 t^{\theta^\prime},
\label{e10}
\end{equation}
where $m_0$ is the initial magnetization and the dynamic exponent
$\theta^\prime$ 
is related to  $x_0$ by
$\theta^\prime =( x_0 -\beta/\nu)/z$
\footnote {In recent literature on
short-time critical dynamics,
 the exponent $\theta^\prime$
is now often denoted as $\theta$. To avoid the confusion with the stiffness 
exponent $\theta$ in spin glasses, however, we recover
the notation $\theta^\prime$ as used in Ref. \cite {jan89}.}. 
For a large variety of systems, the exponent $\theta^\prime$ is
found to be positive, i.e. the magnetization
undergoes an initial increase 
\cite{jan89,li94,sch95,zhe98}. This fact makes
the effect of the initial magnetization $m_0$ very prominent.

Another observable which exhibits the critical exponent $\theta$
is the autocorrelation $A(t)=< L^{-d}\sum_i S_i(0)S_i(t)>$.
Here $S_i(t)$ is a spin variable. At the critical point,
 $A(t)$ evolves with a power law \cite{jan92}
\begin{equation}
A(t)\sim t^{-\lambda},
\quad \lambda={d\over z}-\theta^\prime.
\label{e20}
\end{equation}
Here the initial magnetization is set to $m_0=0$. 
The exponent
$\lambda$ of the autocorrelation still relates to $x_0$
since the exponent $x_0$ is not only the scaling dimension
of the global initial magnetization, but also 
 of the local magnetization density \cite{jan92,zhe98}.

In the past years, the dynamic behavior of 
spin glasses is one of the important subjects
in statistical physics.
The {\it aging} phenomena, i.e. the dependence of dynamic
observables on the temperature and field history of the samples, 
implies that nonequilibrium nature of the 
experimental situation is unavoidable in 
spin glasses \cite{fis91}. 
Therefore, the short-time dynamics
of spin glasses is of special importance. 
The dynamical exponent $\lambda$ of the autocorrelation A(t)
 has been numerically measured by several authors
\cite{hus89,rie93,kis96}.
Up to now, however,
the critical initial increase of 
the order parameter
has not been investigated and correspondingly the exponent $\theta ^ \prime$
has not been obtained directly.
The difficulty here comes from the fact that
the standard order parameter in spin glasses is rather different from
that of simple spin systems with a second order
phase transition.

In this letter we report our novel approach to the
short-time dynamics of the  three-dimensional
$\pm J$ Ising spin glass. For the first time, we measure the
new exponent $\theta ^ \prime$. Using the scaling relation
$\lambda={d / z}-{\theta^ \prime}$ and taking the $\lambda$
from the recent reference \cite{rie93} as input, 
we estimate the dynamic exponent $z$.

The system under consideration is the 3D Edwards-Anderson model
described by the Hamiltonian
\begin{equation}
H=-\sum_{<ij>} J_{ij} S_i S_j
\label{e30}
\end{equation}
where the spins $S_i=\pm 1$ are located on the sites of a $L^3$
cubic lattice with periodic boundary condition
and the sum is over the nearest neighbors.
The interaction $J_{ij}$ take randomly $+1$ or $-1$ 
with probability $1/2$.

For spin glasses, the measurement of the new  
exponent $\theta ^ \prime$ is not straightforward.
Here the magnetization is not the order parameter
 and does not present the behavior 
of Eq. (\ref {e10}). 
The standard order parameter of spin glasses
is a kind of quenched average of the square
of the local magnetization \cite {edw75}. 
It is not clear what would be the corresponding
behavior of Eq. (\ref {e10}).

Recently, a pseudo-magnetization has been introduced,
which may be a candidate of an alternative order parameter
of spin glasses. The pseudo-magnetization is rather similar
to the standard magnetization in simple spin systems,
e.g. the Ising model.
The general idea is that if we
can find out a ground state of the system
(for fixed couplings), then
we define the projection of the spin configurations
onto the ground state as
the pseudo-magnetization
\begin{equation}
m(t)={1\over L^d} \sum_i S_i(t) S_i ^0,
\label{e40}
\end{equation}
where $\{S_i^0\}$ denotes the ground state configuration.
In case of the Ising model, a ground state is a configuration
with all spins in a same direction,
i.e., the pseudo magnetization  is just the standard magnetization.
The pseudo magnetization has recently been applied
to the two-dimensional spin glass to determine
the critical exponents.
The results are promising \cite{rie96}. 
Therefore, we expect that starting from a random initial state
 with small 
pseudo-magnetization $m_0$, the dynamic 
evolution of this seudo magnetization will obey
the power law in Eq. (\ref {e10}).

In this paper, we are interested in three dimensional spin glasses.
However, for a three-dimensional spin glass,
the search for the ground state is NP-Hard, i.e. there is no
polynomial algorithm for it. For systems with 
relatively large lattice size, it is not possible to find out the 
exact ground states. Instead, we can only obtain approximate
ground states. 
This approximation could cause severe correction to the dynamic scaling
behavior of $m(t)$. To reduce this effect 
we introduce the technique of replicas.
For a sample distribution $\{J_{ij}\}$,
we first find out an approximate ground state,
then define two replicas by starting from
two independent initial configurations, but 
both with a small initial pseudo-magnetization $m_0$. 
We update these two replicas with the Metropolis algorithm, 
and measure the dynamic evolution of the overlap 
\begin{equation}
q(t)={1\over L^d} \sum_i S_i^1(t) S_i ^2(t).
\label{e50}
\end{equation}
Here $ S_i^1$ and $ S_i^2$ denote the spins for the two replicas.
If the projection 
of the spin configuration of each replica on the ground state
increases with a power law in Eq. (\ref {e10}),
it is a natural assumption that the mutual 
projection $q(t)$ of these two configurations  should increase
with a power law
\begin{equation}
q(t)\sim m_0^2 \ t^{2\theta^ \prime}.
\label{e60}
\end{equation}

\begin{figure}[p]\centering
\epsfysize=12cm
\epsfclipoff
\fboxsep=0pt
\setlength{\unitlength}{1cm}
\begin{picture}(13.6,12)(0,0)
\put(0,0){{\epsffile{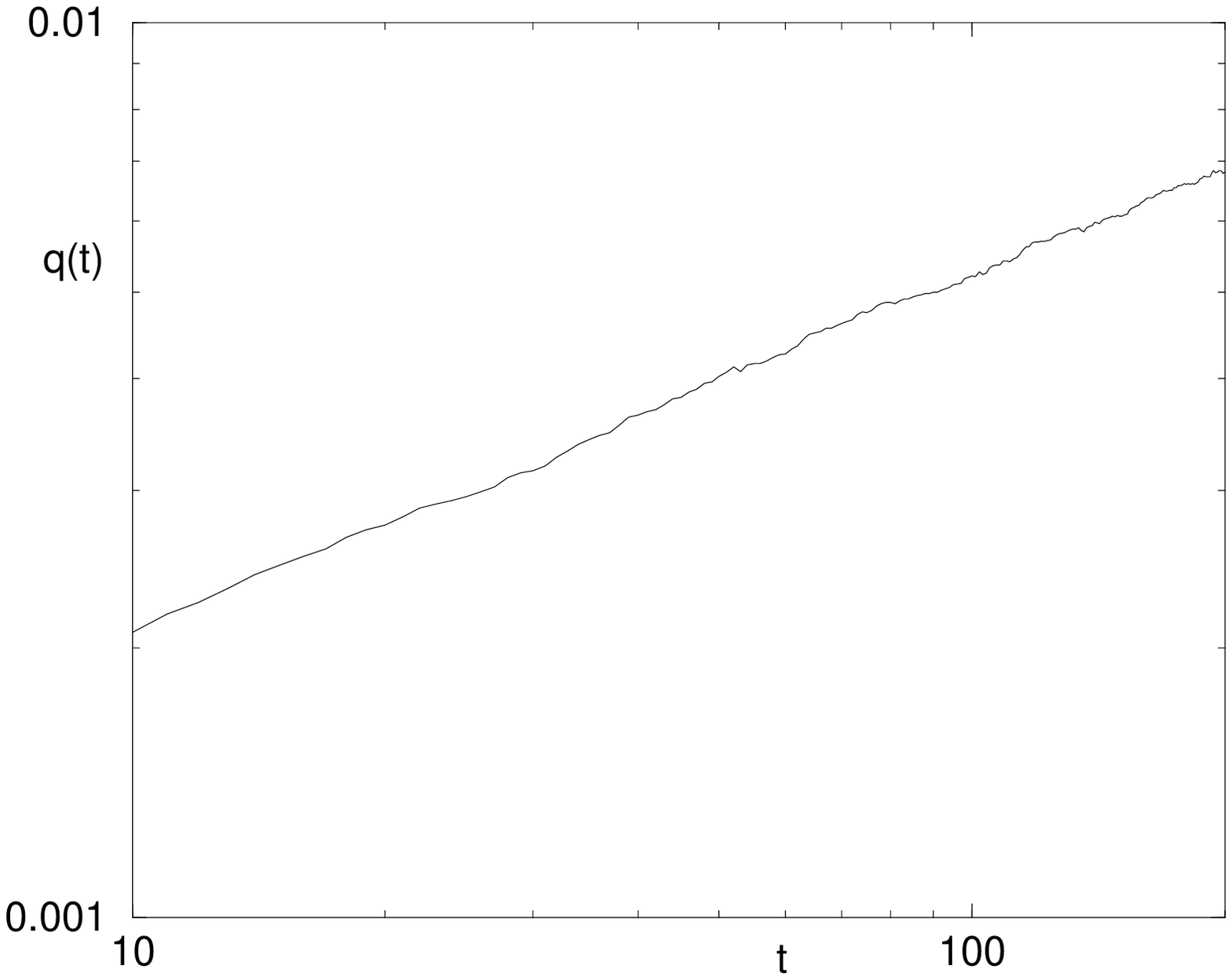}}}
\end{picture}
\caption{ The time evolution of the $q(t)$ for 2D Ising 
Model at $T_c$ is plotted in double-log scale.
The lattice size is set as  $L=128$ and the initial magnetization 
$m_0$ for each replica is $0.02$.
}
\label{f1}
\end{figure}

In order to verify this assumption, we have performed 
simulations for the two-dimensional Ising model.
Starting from two replicas with 
independent initial configurations, both with a small initial 
magnetization $m_0=0.02$ on the lattice $L=128$, we update
these replicas at the critical temperature
with Metropolis algorithm until 200
Monte Carlo steps, and measure the dynamic evolution of the 
overlap $q(t)$.
 Average was taken over $200000$ couples
of independent initial configurations.
In Fig.~\ref {f1}, the time evolution of $q(t)$ is plotted on
double-log scale. After about 10 Monte Carlo steps, 
the curve shows a good power law behavior. 
From the slope in the time interval $[10,200]$ we estimate 
 $2 \theta^ \prime=0.391(4)$. Here the error is estimated
by dividing the samples into three 
subsamples. This result agrees well with the direct measurement
of $\theta^ \prime$ for the Metropolis algorithm
 reported in Ref. \cite{oka97a}, 
$\theta^ \prime =0.197(1)$. 

To  proceed the simulations for the spin glass we need to
search for an approximate ground state
for each sample distribution  $\{J_{ij}\}$.
We employ a simple and efficient algorithm proposed in Ref. \cite{che98}. 
On a lattice $L=16$,  it takes
about 1000 second on a 400 MHz ALPHA station to obtain
an average energy per site $e=-1.7829(3)$, which should be good
enough by referring to the estimated ground state energy
$e_\infty =-1.785708(75)$
for the 3D $\pm J$ spin glass \cite{pal96}. 
With the ground states $\{S_i^0\}$ at hand,
we generate an initial configuration by 
randomly setting the $S_i$ to be  $+S_i^0$ or $-S_i^0$ 
with the probability $(1+m_0)/2$ or $(1-m_0)/2$, respectively.
The configuration generated in this way has an averaged 
initial pseudo magnetization $m_0$. 

Rigorously speaking, the critical exponent $\theta^\prime$ 
is defined in the 
limit $m_0\rightarrow 0$. This requires the
initial pseudo magnetization $m_0$ to be small enough.
We have performed the simulations on 
a lattice $L=16$ with $m_0=0.06, 0.04$ and
$0.02$, respectively.
The results show that there is already no 'finite $m_0$ effect'
for $m_0=0.02$. Besides, by analyzing rough results on
lattices $L=8$ and $L=12$ we find that on the lattice $L=16$  
the finite size effect can be ignored.
Therefore, in our simulations we mainly 
set $m_0=0.02$ and $L=16$.

In recent references,
the critical temperature for the 3D $\pm J$ Ising spin glass 
is usually cited as $T_c=1.175(25)$ \cite{ogi85a,bha88}.
 We perform our simulations with three different temperatures
around this $T_c$, i.e at $T=1.05$, $T=1.175$ and $T=1.3$.
For each sample of $\{J_{ij}\}$
we take an average over $10000$ 
initial configurations, and the final result 
is averaged over 1000 samples of the $\{J_{ij}\}$. 
In each run, we update
the replicas until $500$ Monte Carlo steps. 

\begin{figure}[p]\centering
\epsfysize=12cm
\epsfclipoff
\fboxsep=0pt
\setlength{\unitlength}{1cm}
\begin{picture}(13.6,12)(0,0)
\put(0,0){{\epsffile{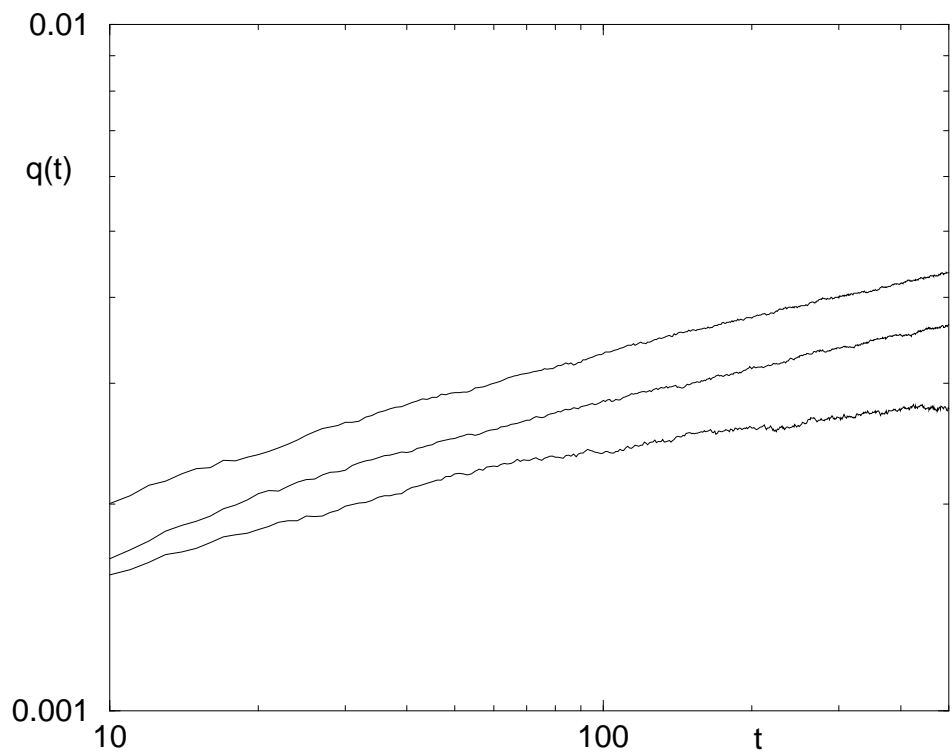}}}
\end{picture}
\caption{ The time evolution of the $q(t)$ for 3D $\pm$ Ising
spin glass is plotted in double-log scale.
From the top to the bottom, the temperatures are $T=1.05$,
$1.175$ and $1.3$ respectively. 
}
\label{f2a}
\end{figure}

In Fig.~~\ref {f2a}, the time 
evolution of $q(t)$ for the three temperatures is 
plotted on log-log scale. At $T_c=1.175$, $q(t)$ shows
a power law increase after about $30$ Monte Carlo steps.
From the slope in the time interval $[30,500]$, 
we obtain the exponent $2 \theta ^ \prime=0.17(1)$, i.e.
the exponent $\theta ^ \prime=0.085(5)$. 
At $T=1.3$, $q(t)$ obviously deviates from the power law
behavior. This is a typical
behavior in a second order phase transition for $T > T_c$.
At $T=1.05$, $q(t)$ seems to show still a power law with 
a similar exponent $2 \theta ^ \prime=0.18(1)$. 
This could mean that the critical temperature is actually
between $1.05$ and $1.175$. Indeed,
in a recent paper \cite{kaw96}, $T_c$
is measured as $1.11(4)$.
In principle, with extensive simulations to a longer time,
the critical temperature $T_c$ can be located
from the short-time critical behavior   \cite{sch96,zhe98}.
However, this requires simulations on lager lattices
 and is beyond our present power of computing.

With the critical exponent $\theta ^ \prime$ in hand, 
we can test the scaling relation $\lambda = d/z - \theta ^ \prime$.
In Ref. \cite{rie93},
$\lambda$ has been estimated to be $3.9(1)$. This value
agrees very well with the experiment \cite{gra87}.
Takinging $\lambda$ and $\theta ^ \prime$ as input,
we obtain the dynamic exponent $z=6.3(2)$.
This value is in agreement with the results $z=6.1(3)$ in \cite{ogi85a},
$z=5.85(30)$ in \cite{blu92} and $z=6.0(5)$ in \cite{bha92}.

In conclusion, we have numerically simulated the dynamic relaxation
process of the 3D $\pm J$ Ising spin glass starting from
random initial states with small pseudo-magnetization $m_0$.
The initial power law increase of the overlap $q(t)$ 
between two replicas  
is observed and the critical exponent $ \theta ^ \prime$ is
measured. The scaling relation  $\lambda = d/z - \theta ^ \prime$
is confirmed.

{\bf Acknowledgements}:
Work supported in part by the Deutsche 
Forschungsgemeinschaft;
Schu 95/9-1 and SFB~418.

\end{document}